\documentclass[aps,twocolumn,superscriptaddress,floatfix]{revtex4}
\usepackage{epsfig}
\usepackage{graphicx}
\usepackage{amssymb}
\usepackage{subfigure}
\usepackage{color}

\begin{document}
\title{Nucleation barriers in tetrahedral liquids spanning glassy and crystallizing regimes}

\author{Ivan Saika-Voivod}

\affiliation{Department of Physics and Physical Oceanography,
Memorial University of Newfoundland, St. John's, NL, A1B 3X7, Canada}

\author{Flavio Romano}
\affiliation{Dipartimento di Fisica, Sapienza -- Universit{\`a} di Roma,
P.le A. Moro 5, 00185 Roma, Italy}

\author{Francesco Sciortino}
\affiliation{Dipartimento di Fisica and CNR--ISC, Sapienza -- Universit{\`a} di Roma, P.le A. Moro 5, 00185 Roma, Italy}

\date{\today}

\begin{abstract}
Crystallization and vitrification of tetrahedral liquids are important both
from a fundamental and a technological point of view. Here, we study via
extensive umbrella sampling Monte Carlo computer simulations the nucleation
barriers for a simple model for tetrahedral patchy particles in the regime
where open tetrahedral crystal structures (namely cubic and hexagonal
diamond and their stacking hybrids) are thermodynamically stable. We show
that by changing the angular bond width, it is possible to move from a
glass-forming model to a readily crystallizing model. From the shape of
the barrier we infer the role of surface tension in the formation of the
crystalline clusters. Studying the trends of the nucleation barriers with
the temperature and the patch width, we are able to identify an
optimal value of the patch size that leads to easy nucleation.
Finally, we
find that the nucleation barrier is the same, within our numerical
precision, for both diamond crystals and for their stacking
forms.
\end{abstract}

\maketitle
\section{Introduction}

Crystallization is central to several fields, from materials research to
biological science, not to mention its technological
relevance~\cite{kashchiev,keltonbook}. Several human pathologies are also
caused by crystal nucleation in protein solutions~\cite{vekilovsoftmatter}.
Understanding crystal nucleation requires both the evaluation of the
stability fields of the fluid and crystal phases (i.e. knowledge of the
chemical potentials of all possible phases), as well as the evaluation of
the thermodynamic barriers controlling the formation of the stable phase
and of the kinetic pre-factors fixing the timescale of the diffusional
processes. 

In recent years, several numerical methodologies have been developed for
accurately evaluating phase diagrams from the free energies of the fluid
and crystal phases. We refer the reader to the review by Vega and
coworkers~\cite{VegaJPCM}. Also for crystallization, various methods are
now available for calculating free energy barriers and nucleation
rates~\cite{frenkel_1996, frenkel_1999, frenkel_2001, frenkel_2004,
frenkel_2005}, making it possible to generate accurate data for model
potentials and, more importantly, to compare the numerical results with
theoretical predictions, mostly based on classical nucleation theory
(CNT)~\cite{gibbs, volmer, farkas, becker,kelton,pablo}, as well as, when
possible, with experimental data.

One of the main motivations for the development of the special methods for
studying nucleation is the proper sampling of the equilibrium cluster size
distribution $N(n)$ within the metastable liquid, i.e. the number of
crystal-like clusters composed of $n$ particles. The work of forming a
cluster of size $n$ from the liquid is given in terms of $N(n)$
by~\cite{frenkel_2004},
\begin{equation}
\label{NnCNT}
\beta \Delta G(n) = -\ln \left[ \frac{N(n)}{N_p}\right]\:,
\end{equation}
where $N_p$ is the
number of particles in the system and $\beta \equiv (k_{\rm B}T)^{-1}$, where $T$ is
the temperature and $k_{\rm B}$ is the Boltzmann constant.
Below the melting temperature, $\beta \Delta G(n)$ has a maximum value
$\beta \Delta G^*$ at a critical nucleus size $n^*$. $\beta \Delta G^*$
enters into the expression for the rate of nucleation exponentially, and is
therefore the main consideration in determining the rate from a
thermodynamic perspective. Very generally, the larger the difference in
chemical potential between liquid and crystal $\Delta \mu$, often referred
to as the driving force for crystallization, the smaller $\beta \Delta G^*$
is. Conversely, the surface tension $\gamma$ between crystallite and
surrounding liquid always acts against the growth of crystallites, and so
$\beta \Delta G^*$ increases with $\gamma$. Hence, the ability of a system
to crystallize is governed by the interplay between $\Delta \mu$ and
$\gamma$.

Recently, simulation studies have begun to address crystallization of
tetrahedral liquids~\cite{md1,md2,md4,vegaprl,mousseau,Noya2010}. This class of
interesting liquids includes biological and technologically important
molecular and atomic materials such as water, silica, silicon and carbon.
One common feature in these systems is the formation of open, low density
structures such as the diamond cubic (DC) crystal. For carbon, simulations
have shown the importance of the liquid structure in governing nucleation
barriers~\cite{valerianiCarbon,valerianiCarbon2}. In a generalized form of the Stillinger-Weber model for silicon~\cite{SW}, 
the
degree of tetrahedrality was shown to have a strong impact on DC nucleation
and its interrelation with the appearance of a metastable liquid--liquid
critical point~\cite{sastryPRL}. Further, for silicon and germanium, the
lower density of the DC crystal with respect to the liquid was shown to
give rise to a preference for nucleation near the surface, with similar
implications for water~\cite{GalliNatMat}. The self-assembly of a DC
colloidal crystals is also of interest in the field of photonics, since
such an ordered structure of dielectric spheres is expected to exhibit a
band structure with a large gap within the visible wavelengths of
light~\cite{natmatmaldovan}.

Tetrahedral liquids are interesting from another perspective as well.
Often, in these systems crystal formation is inhibited by the onset of a
glassy behavior which dramatically slows down the microscopic dynamics,
thus preventing the transition to the ordered structure. The glass forming
ability of silica and the ease with which water crystallizes stand in stark
contrast to each other.

Motivated by these issues of scientific and technological interest,
recently, some of us~\cite{FLAVIOXstal} have thoroughly investigated the
question of how to best design tetrahedral patchy colloidal particles that
spontaneously assemble into open crystal structures through computer
simulations of the Kern-Frenkel (KF) model~\cite{KF}. It was shown that the
angular width of the patch plays a key role in determining if the system,
at low temperatures, forms a disordered glass structure or a crystal, a
result which may also be of interest for interpreting the glass-forming
ability of atomic and molecular systems. When the patch width is small, the
crystal coexists with a fluid and $\beta|\Delta\mu|$ (the driving force for
crystallization) increases quickly with undercooling, while when the patch
width is large, the crystal coexists with a fully bonded liquid state and
$\beta |\Delta \mu|$ grows only moderately with undercooling. It was also
shown that the model spontaneously forms a crystal composed of a mixture of
two open tetrahedral structures, the DC and the diamond hexagonal (DH)
crystals, and that the chemical potentials for the two polymorphs were
indistinguishable within the precision of the calculations.

In this article we proceed one step further by investigating, for the same
tetrahedral model, the patch width dependence of the free energy barrier to
nucleating the DC and/or DH crystals. Interestingly, we find that, at
comparable undercooling, the barrier height $\beta \Delta G^*$ does not
monotonically decrease with increasing patch width as one would expect
from a consideration of $\beta |\Delta \mu|$ alone. Indeed, comparing the
barrier shape with CNT predictions, we find that the surface tension
$\gamma$ increases on decreasing patch width, a result that we tentatively
connect to the larger structural and density difference between fluid
and crystal. We do find that for narrow patches $\beta \Delta G^*$
generally shows a strong $T$ dependence, rapidly decreasing to a homogenous
nucleation limit, while for wide patches $\beta \Delta G^*$ decreases more
slowly with decreasing $T$, allowing the system to reach the glassy regime.
Additionally, at the widest patch studied, the barriers 
remain very large (of
the order of 50 $k_BT$) even for
significant undercooling. This confirms that despite the benefit of a
lower surface tension, the lack of a buildup of a large difference in
chemical potential is mainly responsible for disfavoring crystallization of
particles with wide patches.
The presented results confirm that DC/DH will
spontaneously form when the angular width of the patch is sufficiently
small.

The remainder of this article is organized as follows. In Section II, we
outline our criteria for defining crystal-like clusters and the umbrella
sampling Monte Carlo procedure for calculating nucleation barriers. In
Section III we present our results, including a fairly detailed examination
of the robustness of $\beta \Delta G^*$ to varying the cluster-defining
criteria and the dependence of $n^*$ and cluster composition (e.g.
percentage of DC particles) on them. We then present our summary and conclusions
in Section IV.

\section{Methods}

We perform biased Monte Carlo (MC) simulations~\cite{frenkel_book} at
constant $T$ and pressure $P$ of the Kern-Frenkel model~\cite{KF} with
$N_p=1000$ particles, each with hard sphere diameter $\sigma$, and each
having four tetrahedrally arranged attractive patches of range $\delta =
0.24 \sigma$, strength $u_0$ and angular width
$2\theta$~\cite{FLAVIOXstal,FLAVIOPD}. We report $T$ and $P$ as
dimensionless quantities, after rescaling by $u_0/k_B$ and $u_0/\sigma^3$,
respectively. We investigate four different values of $\theta$, namely
those corresponding to 
$\cos{\theta} = 0.98$, 0.96, 0.94 and 0.92 ($\theta = 11.5^\circ$,
16.3$^\circ$, 19.9$^\circ$ and 23.1$^\circ$).  For the value of $\delta$ we use,
the condition $\cos{\theta} > 0.9151$ guarantees that a patch can only 
accommodate a single bond, and therefore the maximum number of bonds per particle
is restricted to four~\cite{FLAVIOPD}.


\begin{figure}[tb]
\hbox to \hsize{\epsfxsize=1\hsize\epsfbox{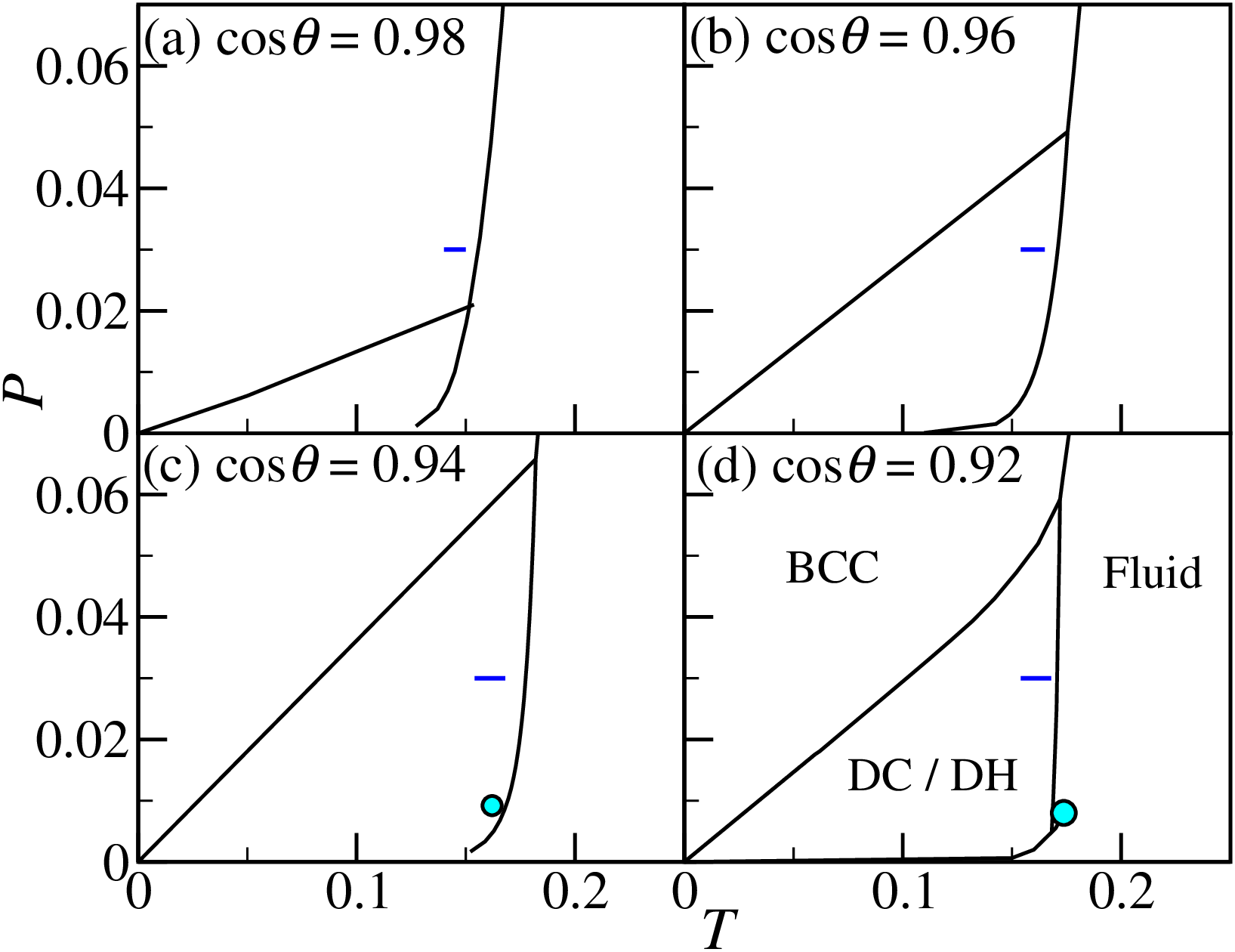}}
\caption{
Equilibrium phase diagrams for all the models studied in the low $P$, low
$T$ region, where the open crystal structures are stable. The short blue
lines correspond to the temperatures for which the nucleation barriers have
been investigated. The circles mark the gas--liquid critical point. For
the narrow patch models [(a) and (b)], nucleation is so effective that the
location of the critical point can not be properly estimated. However, we
have checked that the studied isobar is located significantly above the
critical pressure.
}
\label{fig:phds-barriers}
\end{figure}

Fig.~\ref{fig:phds-barriers} shows the phase diagrams for the Kern-Frenkel
models we study here, obtained either directly from Ref.~\cite{FLAVIOXstal}
[panels (a) and (d)] or by extending the calculations to cases not reported
in that work. The phase diagrams show high and low density fully bonded
crystal phases, in which each particle is bonded with four neighbors (i.e.
the energy per particle is $-2u_0$). The low density crystal is an
open tetrahedral
structure and it can exist in two polymorphs, DC and DH. As noted in
Ref.~\cite{FLAVIOXstal}, the chemical potentials for the DH and DC
structures are largely indistinguishable, and indeed it has been observed
that when the liquid crystallizes spontaneously to the open tetrahedral
structure, it does so by forming a stacking of DC and DH planes~\cite{FLAVIOXstal}.
Indeed, DH and DC crystalline layers can stack in an analogous way to the
FCC/HCP stacking in hard-spheres. The dense phase, composed of two
interpenetrating fully bonded DC structures, is a body centered cubic
(BCC) crystal. The fluid separates into gas and liquid phases below the
critical temperature. Since the range of the potential is short compared to
the particle size, the critical point is almost always metastable with
respect to the crystal phase. 
 
In the following, we study the crystallization barriers to the DC/DH
crystal at one selected pressure. Specifically, we choose $P=0.03$ in order
to avoid interference with gas-liquid phase separation.
For all the models studied, with the exception of $\cos\theta=0.98$,
the most stable phase at that pressure is the open tetrahedral crystal.
In the case of $\cos\theta = 0.98$, BCC is the stable phase at this pressure,
but as previous studies have shown, spontaneous crystallization results
always in the open structure~\cite{FLAVIOXstal}, an example of the 
Ostwald step rule~\cite{OSTWALD1,OSTWALD2,OSTWALD3}. 
 
The nucleation to tetrahedral crystals has been studied previously in
simulations~\cite{valerianiCarbon,valerianiCarbon2,GalliNatMat} and we follow the
established methodology to evaluate the free energy barriers. A novel
issue arises from the presence of two crystals with different symmetries
in the same stability field, as we discuss in the following.
 
\begin{figure}[tb]
\hbox to \hsize{\epsfxsize=1\hsize\epsfbox{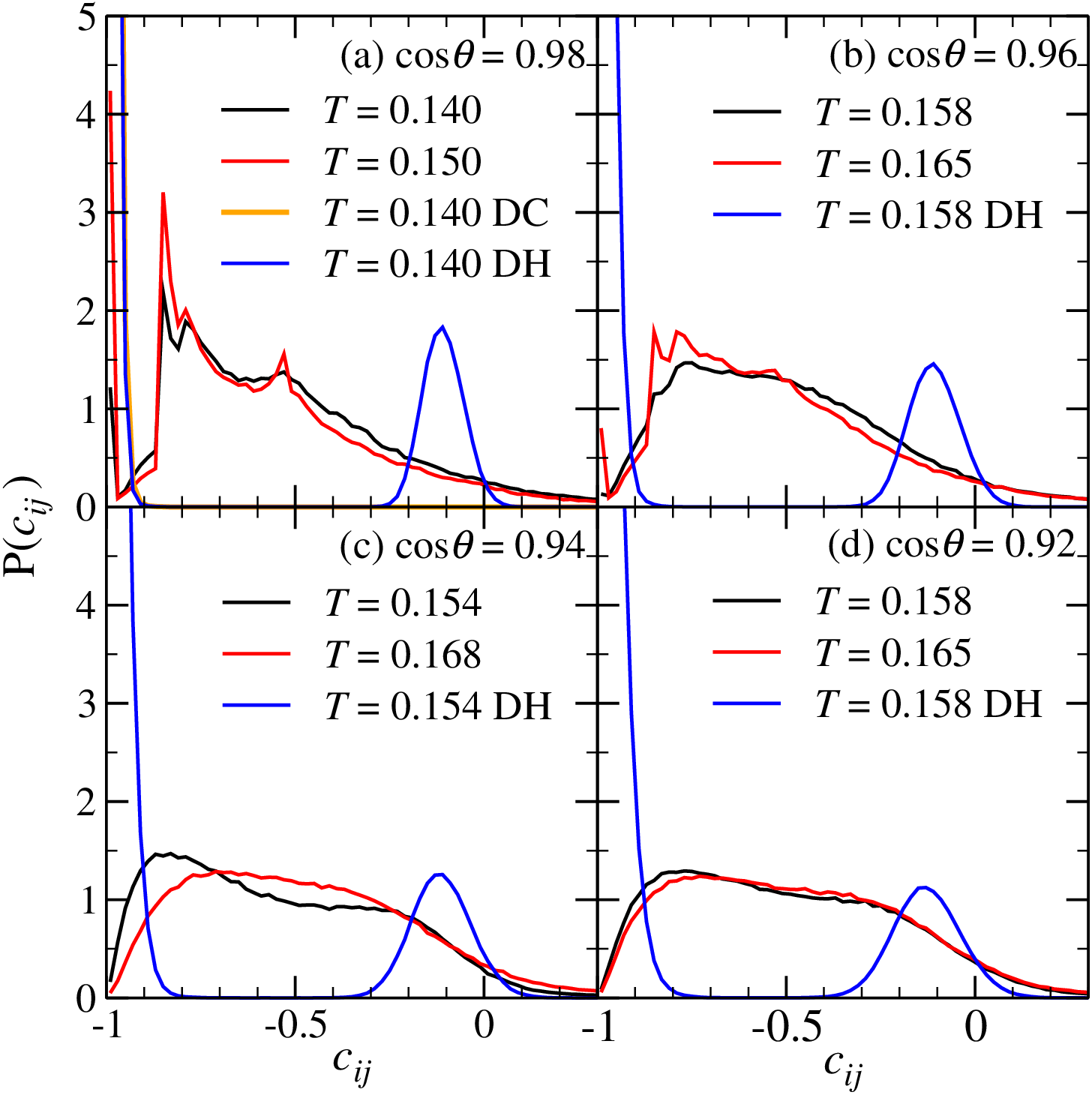}}
\caption{\label{fig:cij}
Probability distributions of the dot product $c_{ij}$ in the liquid and
crystal structures. The distribution for DC, shown only in panel (a), is a
single peak near -1 and is nearly indistinguishable from the DH peak at -1
on the scale of the plots. The distributions for the crystals do not vary
significantly over our $T$ range.
}
\end{figure}

We use Steinhardt bond order parameters~\cite{steinhardt} based on spherical harmonics of order $l=3$.
For each particle we define the complex vector
\begin{equation}
q_{lm}(i)=\frac{1}{N_{\rm b}(i)}\sum_{j=1}^{N_b(i)}Y_{lm}(\hat{r}_{ij}),
\end{equation}
where the sum is over the $N_{\rm b}(i)$ neighbors of particle $i$, defined
as those particles within a distance of $(1+\delta)\sigma=1.24\sigma$ of
particle $i$. The dot product
\begin{equation}
c_{ij} = \sum_{m=-l}^l \hat{q}_{lm}(i) \hat{q}^*_{lm}(j),
\end{equation}
where
\begin{equation}
\hat{q}_{lm}(i) = q_{lm}(i)/\left(\sum_{m=-l}^l \left| q_{lm}(i)\right|^2  \right)^{1/2}
\end{equation}
and
$
\hat{q}^*_{lm}(i)
$
is its complex conjugate, determines the degree of orientational
correlation between neighboring particles $i$ and $j$. Fig.~\ref{fig:cij}
shows the distributions for the fluid and the DC and DH crystals. The DC
distribution is peaked around $c_{ij}=-1$ only, while the DH crystal also
shows a peak near $c_{ij}=-0.1$. In the DH crystal, each particle has three
neighbors with $c_{ij} \approx -1$ and one with $c_{ij} \approx -0.1$,
while in the DC crystal all four bonded neighbors have $c_{ij} \approx -1$.
This provides a local basis for distinguishing particles as being  
DC or DH. The
fluid distributions are very wide and show sharp peaks for large values of
$\cos \theta$ (small bonding angles). The peaks become more intense on
heating, which we attribute to a lower density and higher energy; we
identify the peaks as signals of specific geometrical assemblies in the
fluid with unfilled bonds. For example, a dimer has $c_{ij}=-1$, while both
bonds in a 
trimer have $c_{ij}=-0.82$
 
The possibility of separating the $c_{ij}$ ranges in which crystal-like or
fluid-like particles are mostly contributing offers a way of associating a
value of $c_{ij}$ with a local structure (crystal or fluid). Usually a
threshold number of crystal-like connections is selected to distinguish
particles as being fluid-like or crystal-like. In the following, we start
by defining a crystal-like connection between neighbors as one with
$c_{ij}<q_u$, with $q_u=-0.87$ and a crystal-like particle as one which has
three or more crystal-like connections. This definition does not allow one
to discriminate between DC and DH and hence appears to be a reasonable
first step in the investigation of the nucleation
barriers~\cite{valerianiCarbon,valerianiCarbon2,GalliNatMat}. We complement this study with
an additional investigation where we differentiate between DC and DH by
requiring that a solid-like particle have four neighbors with
$c_{ij}<-0.87$ in the DC case or three neighbors with $c_{ij}<-0.87$ and
one with $-0.3< c_{ij}<0.1$ in the DH case. For completeness, we probe
three cases: (i) the case where the growing crystal is composed only of DC
particles; (ii) the case where the growing crystal is composed only of DH
particles; (iii) the case where the growing crystal is composed of DC or DH
particles. 

We follow the standard methodology for defining a biasing potential which
helps the formation of crystalline clusters. To this aim we add to the
Kern-Frenkel potential a perturbation given by
\begin{equation}
\phi = \kappa (n_{\rm max} - n_0)^2,
\end{equation}
where $\kappa$ is a suitably chosen constant that controls the range of
sampled crystal cluster sizes, centered near $n_0$, and
$n_{\rm max}$ is the size of the largest crystalline cluster in the system.
Depending on the steepness of $\Delta G(n)$, we adjust the value of
$\kappa$ to obtain good sampling, but for almost all simulations, $\kappa =
0.075$. To define a crystal-like cluster, we state that two neighboring crystal-like
particles are part of the same cluster.

New configurations in the umbrella sampling (US) MC chains are generated as
follows. Given a starting configuration with largest cluster $n^0_{\rm
max}$, we perform a trajectory of 20 Metropolis MC steps, where one such
step is on average $N_{\rm p} - 1$ particle translations and rotations and
one volume change, in order to arrive at a configuration with largest
cluster $n^1_{\rm max}$. The new configuration is accepted with probability
$\min({1, \exp{\{-\beta [ \phi(n^1_{\rm max}) - \phi (n^0_{\rm max})
]\}}})$. 
If the new configuration is accepted, it becomes the starting point for the next MC trajectory.
If it is rejected, the entire trajectory is discarded and the $n^0_{\rm max}$ configuration is 
kept and used as the starting point for generating another MC trajectory. 
For the slowest state points, we perform $10^7$ US
attempts. We perform several US simulations for different values of $n_0$.
Generally, two adjacent US windows are spaced by $\Delta n_0=3$ to ensure
good sampling of the reaction path. For each US window we then evaluate the
solid cluster size distribution ${\tilde N}(n)$ and ${\rm P}_{\rm
max}(n)$, the probability that the largest cluster in the system is of size
$n$.

The cluster size distribution $N(n)$ in the $NPT$ ensemble for each window
is worked back from the biased ensemble as in Ref.~\cite{frenkel_2004},
with
\begin{equation}
N(n) = \left < \exp{[\beta \phi(n_{\rm max})]} \tilde N(n) \right >_{\rm biased}
\end{equation}
up to a multiplicative constant, where $\left < \dots \right>$ indicate an
ensemble average. Portions of $\beta \Delta G(n)$ are determined from
simulations at different values of $n_0$ up to additive constants, in
agreement with Eq.~\ref{NnCNT}. These pieces are matched by minimizing the
difference between overlapping portions after discarding data for which
$P_{\rm max}(n)$ is less than 0.01 (to ensure good sampling).
Alternatively, we find the free energy difference $\Delta G (n) - \Delta
G(n-1)$ as a weighted average of the values obtained with simulations at
different $n_0$ in which clusters of size $n$ and $n-1$ have been sampled,
where the weight is given by $\exp[- 2 k (n - n_0)]$. As a result $\Delta
G(n)=\sum_{j=1}^n [\Delta G(j) - \Delta G(j-1)]$. The two procedures give
equivalent results. We then shift the curves so that $\Delta G(0)=0$.
While this is not formally correct, since the number of liquid-like
particles [$N(0)$] is not precisely equal to $N_{\rm p}$, the error is
negligible (of the order of 0.01$\,k_{\rm B}T$ or less). We stress here
that at this stage we do not differentiate between DH and DC particles. We
have checked that indeed the largest crystalline cluster is a mixture of
the two crystal local structures. We will address this point in more detail
later on.

To ensure independence of the initial conditions, we construct our barriers
starting from two different sets of data. In the first set we use the
biasing potential to grow clusters in the fluid, starting the US simulation
at $n_0$ from a configuration extracted from the previous windows. In the
second set, we seed the biased simulations at $\cos\theta=0.98$ with
crystallite-containing configurations from homogeneously nucleating
unbiased MC runs at $\cos{\theta}=0.98$. Simulations at smaller values of
$\cos\theta$ are started from the $\cos{\theta}=0.98$ data set, after
equilibration at higher values of $T$.

\section{Results}

\subsection{Barrier profiles}

\begin{figure}[tb]
\hbox to \hsize{\epsfxsize=1\hsize\epsfbox{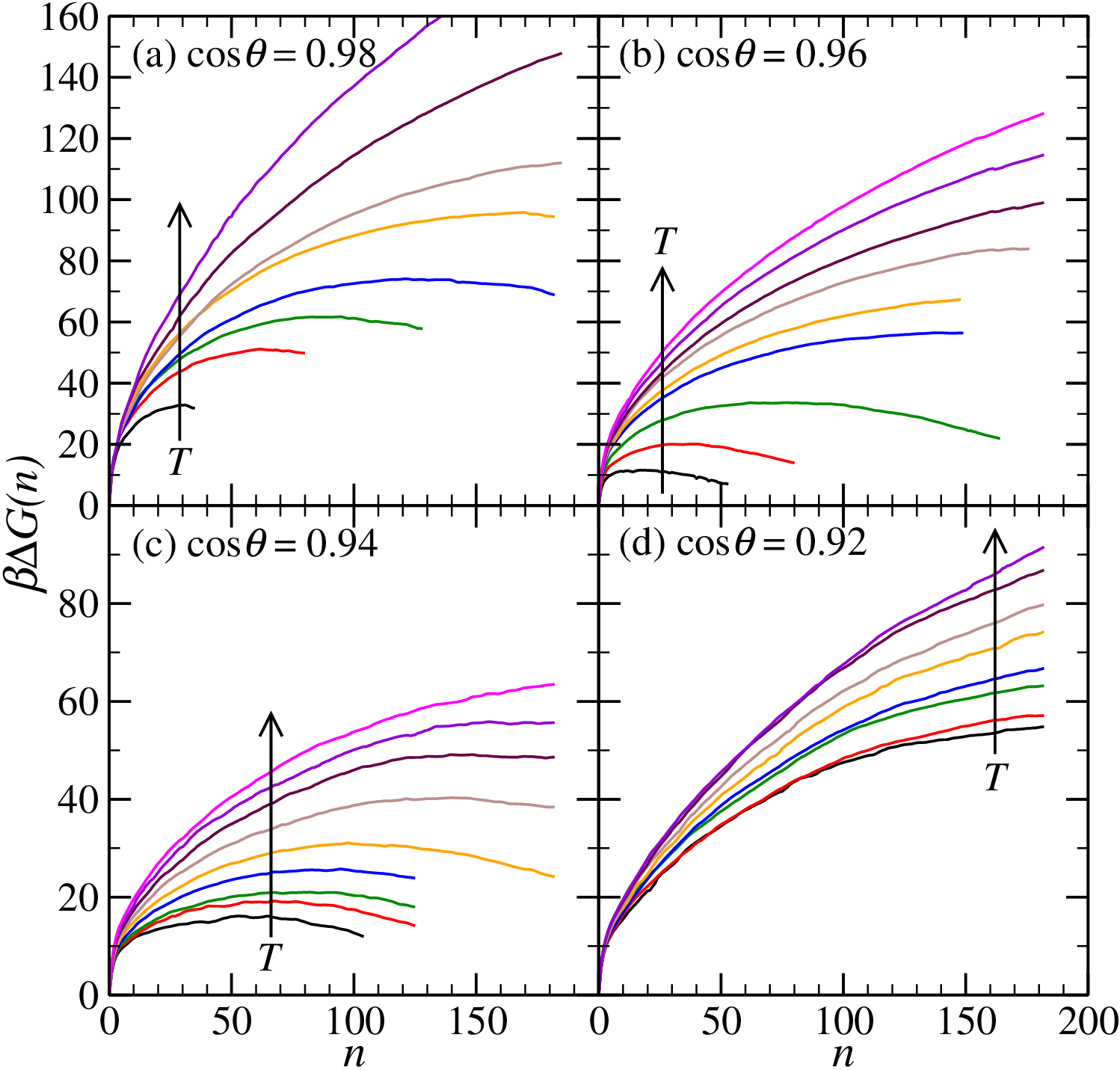}}
\caption{\label{fig:profiles1}
Nucleation barriers $\beta \Delta G(n)$ for various bond angular widths and temperatures.
Panel a) $\cos{\theta}=0.98$, from bottom to top $T = 0.140$, 0.142, 0.143, 0.144, 0.145, 0.146, 0.148, 0.150.
Panel b) $\cos{\theta}=0.96$, from bottom to top $T = 0.154$, 0.156, 0.158, 0.160, 0.161, 0.162, 0.163, 0.164, 0.165.
Panel c) $\cos{\theta}=0.94$, from bottom to top $T = 0.154$, 0.156, 0.158, 0.160, 0.162, 0.164, 0.166, 0.167, 0.168.
Panel d) $\cos{\theta}=0.92$, from bottom to top $T = 0.154$, 0.156, 0.158, 0.160, 0.162, 0.164, 0.166, 0.168.
}
\end{figure}

In Fig.~\ref{fig:profiles1} we show our results for the $\Delta G(n)$
profiles for the four models of different patch widths at various $T$. The
range of $T$ where the barrier can be calculated with the present
methodology is restricted both from above and below for different reasons.
At large $T$ (small supercooling) the critical nucleus is large
compared to the system size.
At low $T$,
different reasons conspire against a proper evaluation of the barrier
height. For example, in the case $\cos{\theta=0.92}$, the slow kinetics
associated with the proximity of a glass transition prevent proper
equilibration already at temperatures where the critical nucleus is of
comparable size to the studied system. In this same case, it has been
suggested on the basis of the explicit evaluation of the difference in
chemical potential between the crystal and the fluid that, at odds with the
standard behavior, barriers do not grow with further supercooling. The
difficulty of evaluating barriers for the wide patch model is
consistent with this proposition. In the case of narrow patches,
$\cos{\theta=0.98}$, we are limited to a barrier height of the order of
30$\,k_{\rm B}T$. For lower $T$, despite slow dynamics not being an issue,
the results of the calculations become more and more affected by the
presence of secondary crystal clusters. While in theory the presence of
more than one cluster is not a problem, in our opinion this effect
highlights an incorrect definition of crystallinity at the local level,
which artificially breaks a single cluster into two (or more) pieces.
Hence, at the lowest $T$, the barrier calculations become difficult owing
to the appearance of clusters of comparable size to the largest one in the
system. A visual inspection of configurations confirms the formation of
secondary clusters occurring next to the primary one, separated by
particles that fall outside the cutoffs defining crystal-like particles. We
recall that for a proper evaluation of the barriers in the low $T$ limit,
where homogeneous nucleation is taking place, one can apply the methodology
based on mean first-passage times~\cite{reguerra, reguerra2, reguerra3,
reguerra4, lundrigan}. Finally, we note that we have checked for state
points where the barrier is of the order of 20$\,k_{\rm B}T$ the quality of
our calculations by comparing the US results with $N(n)$ evaluated in
unconstrained simulations. 

\subsection{Fits}

Within the phenomenological framework of CNT, the work of forming an
$n$-sized cluster can be written as, 
\begin{equation}\label{CNT}
\beta \Delta G(n) = - \beta | \Delta \mu| n + \beta \gamma A,
\end{equation}
where $A$ is the surface area of the cluster. We therefore fit our profiles
to
\begin{equation}\label{eq:CNTfit}
\beta \Delta G(n) = - a\, n + b \,n^{2/3}.
\end{equation}
where $a=\beta | \Delta \mu| $ and $b \sim \beta \gamma$ and we have
assumed a compact cluster. Below, we compare our results for $a$ against
$\beta | \Delta \mu|$, as some studies suggest a very good
correspondence~\cite{frenkel_2004, frenkel_2005, SaikaBKS}, and justify our
assumption that $b$ is equal to $\beta \gamma$ up to a constant factor
which depends on the shape and density of the crystallites. Recently, it
has been shown for a soft-core colloidal model that the procedure we follow
here for determining $n$ may not be sufficient for describing crystal-like
structures, but that the scaling represented by Eq.~\ref{eq:CNTfit} may
still hold, i.e., changing the parameters used to define a crystalline
cluster will change the numerical values of $a$ and $b$~\cite{lechner}. We
explore this in some detail below.

\begin{figure}[tb]
\hbox to \hsize{\epsfxsize=1\hsize\epsfbox{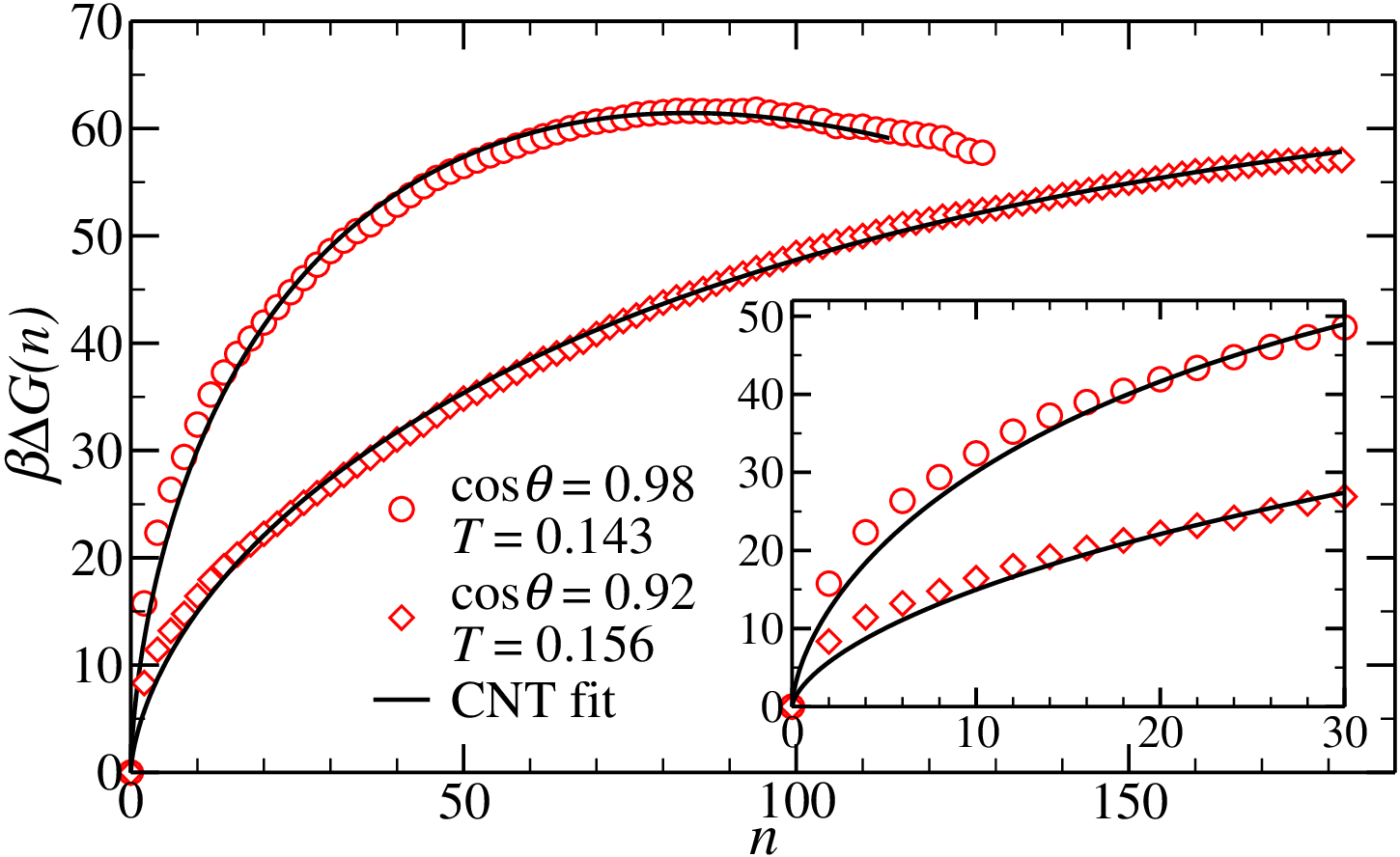}}
\caption{\label{fig:fit}
Fit of two of the barrier profiles studied to the CNT form (Eq.~\ref{eq:CNTfit}). 
While there is a deviation at small $n$, the overall representation of the curve is fairly good,
especially in terms of the barrier height $\Delta G^*$ and size of the critical nucleus $n^*$.
Inset: low-$n$ region of the same data, hilighting the deviation
from the CNT form.
}
\end{figure}

We show two representative $\beta \Delta G(n)$ curves along fits to
Eq.~\ref{eq:CNTfit} in Fig.~\ref{fig:fit}. While there is a deviation
between data and fit at small $n$, the overall description is rather good.
Moreover, the values for $n^*$ and $\Delta G^*$ extracted from the fits
coincide with those obtained directly from the barriers. Improvement of the
small $n$ description could be accomplished along the lines suggested in
Ref.~\cite{ISINGCNT} but it is outside the scope of the present work. 

The fits allow us to plausibly extrapolate our $\Delta G(n)$ profiles to
values of $n$ larger than what we can simulate in our $N_{\rm p}=1000$ system,
allowing us to estimate $n^*$ and $\beta \Delta G^*$ for all the $T$ that
we study. While we do not rely strongly on the accuracy of the
extrapolations, they do provide a stronger sense of the trends in the data.

\begin{figure}[tb]
\hbox to \hsize{\epsfxsize=1.0\hsize\hfil\epsfbox{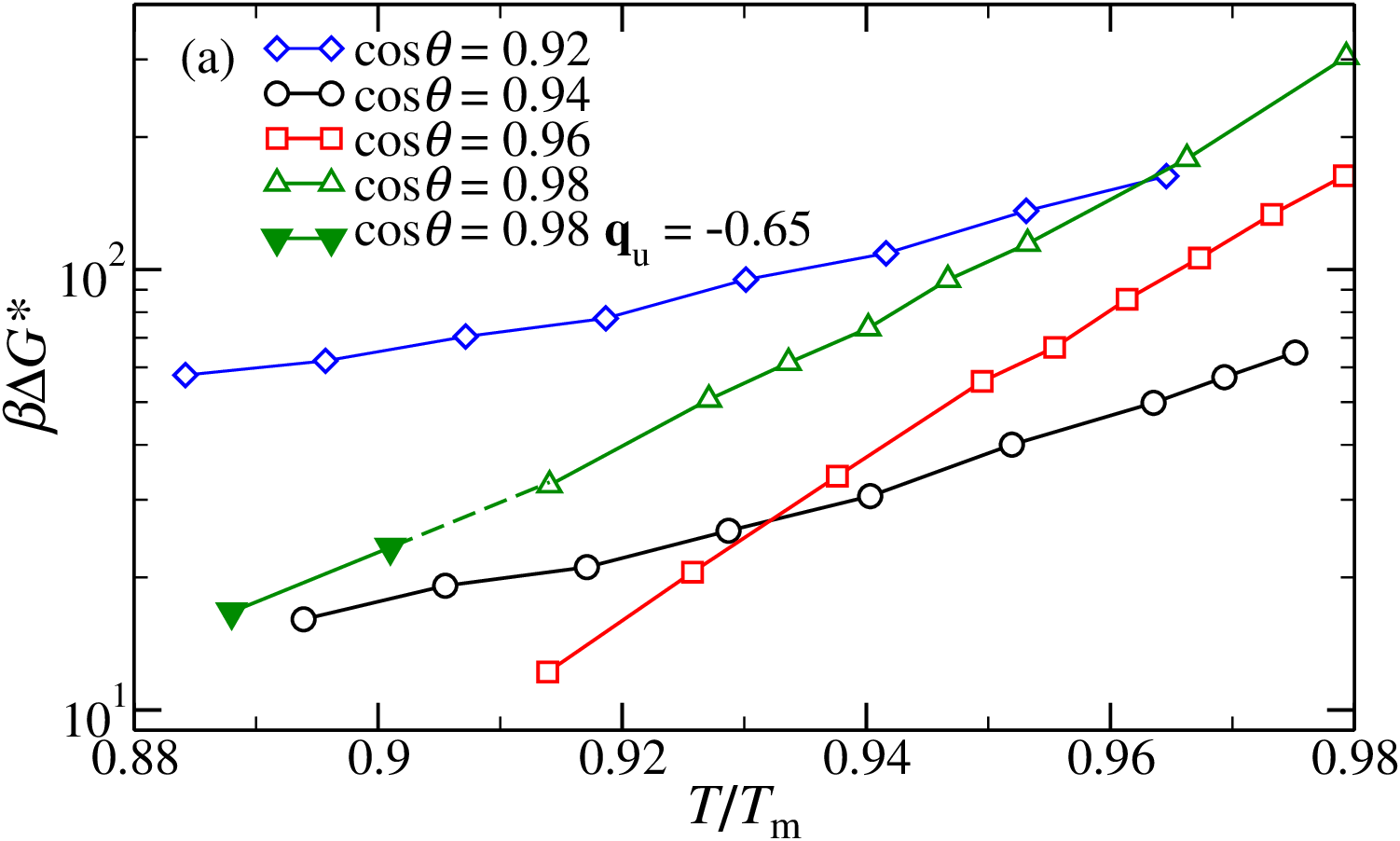}}
\hbox to \hsize{\epsfxsize=1.0\hsize\hfil\epsfbox{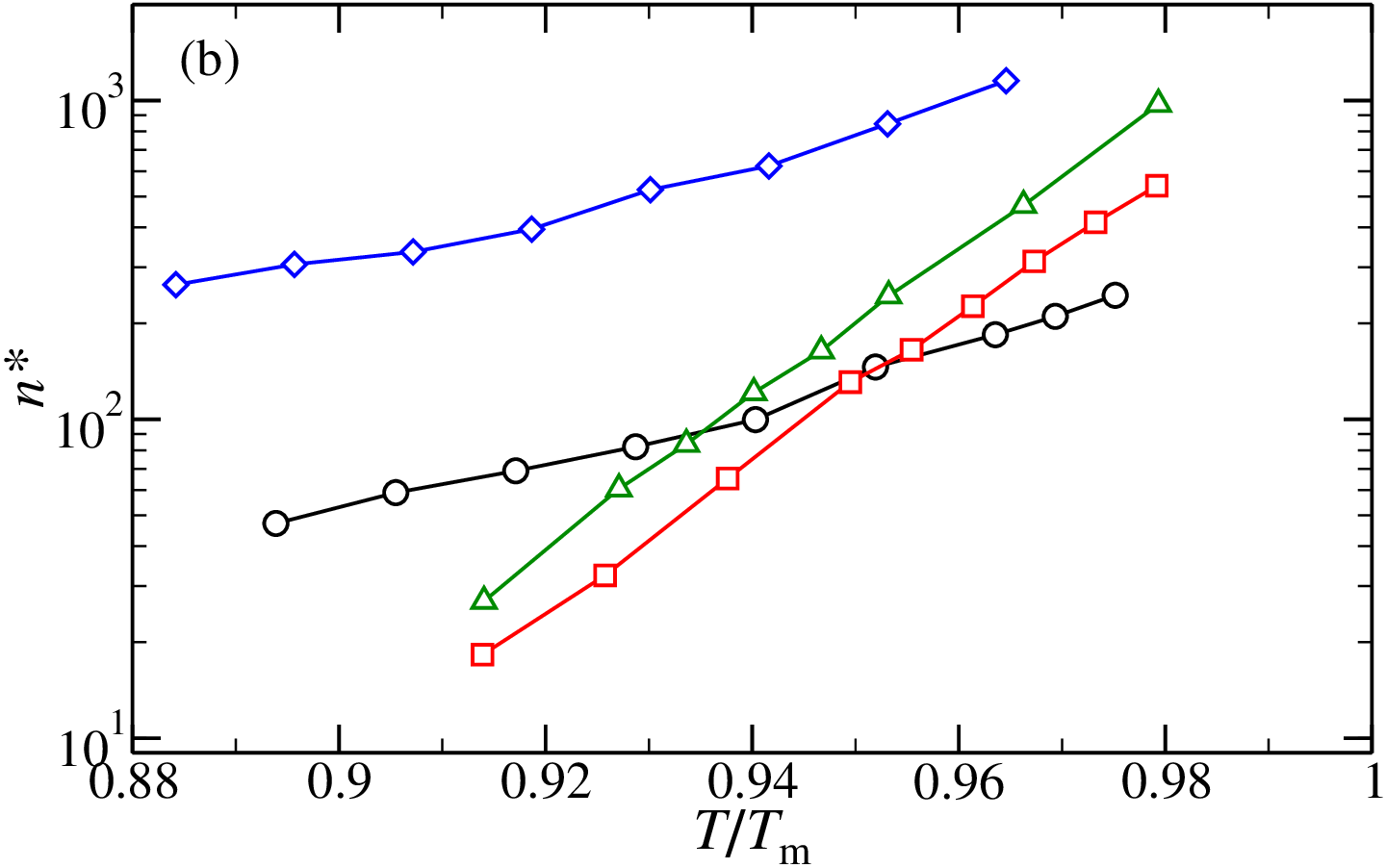}}
\caption{
\label{fig:gstar}
(a) Barrier height $\beta\Delta G ^ *$ as a function of $T/T_{\rm m}$. For
high values of $\cos{\theta}$, the barrier decreases significantly with
decreasing temperature, while the slope is much more shallow for low values
of $\cos{\theta}$. The two points at lowest $T$ for $\cos{\theta}=0.98$
are obtained from US simulations employing a more relaxed $c_{ij}$ cutoff
of $q_u=-0.65$ after checking for consistency at higher $T$.
(b) Critical cluster size $n^*$ as a function of the reduced temperature.
Also in this case the trend with $T$ is stronger in narrow-patch models
than in wide-patch models.
}
\end{figure}

\subsection{$T$ dependence of $\Delta G^*$ and $n^*$}

Fig.~\ref{fig:gstar}(a) shows $\beta \Delta G^*$ as a function of $T/T_m$,
where $T_m$ is the melting temperature for the four models~\cite{melting}
[($\cos{\theta}$, $T_m$): (0.98, 0.153), (0.96, 0.169), (0.94, 0.172),
(0.92, 0.174)]. 
For small angles, the $T$ dependence of $\beta \Delta G^*$ is significant,
and can be modeled rather well in the present range of barriers with an
exponential function. The fast decrease of $\beta \Delta G^*$ provides
evidence for the inevitability of crystallization. For larger angles, we
observe both a significant change in slope, i.e., a much slower decrease of
the barrier height with supercooling, as well as a progressive increase of
the barrier height at fixed supercooling with increasing angle. The trend
is accounted for by the results reported in~\cite{FLAVIOXstal}, namely that
$\beta | \Delta \mu|$ becomes a weaker function of temperature as $\theta$
increases. On the basis of the results presented in
Fig.~\ref{fig:gstar}(a), the model with $\cos{\theta}=0.96$ appears to be
the optimal candidate for crystallization from a thermodynamic perspective,
since modest supercooling is sufficient to induce barriers of
height of the order of 10$\,k_BT$. 

Fig.~\ref{fig:gstar}(b) shows $n^*$ as a function of $T/T_m$, showing
similar trends as for $\beta \Delta G^*(T)$. The $\cos{\theta}=0.98$ and
0.96 models are similar, attaining small $n^*$ for a relatively small
degree of supercooling, while the models with wider patches seem to require
larger sizes of critical nuclei as $T$ decreases. The $T$ dependence of
$n^*$ is much flatter, suggesting that the critical nucleus remains
significantly large even under deep supercooling.

\subsection{Surface tension}

\begin{figure}[tb]
\hbox to \hsize{\epsfxsize=1\hsize\epsfbox{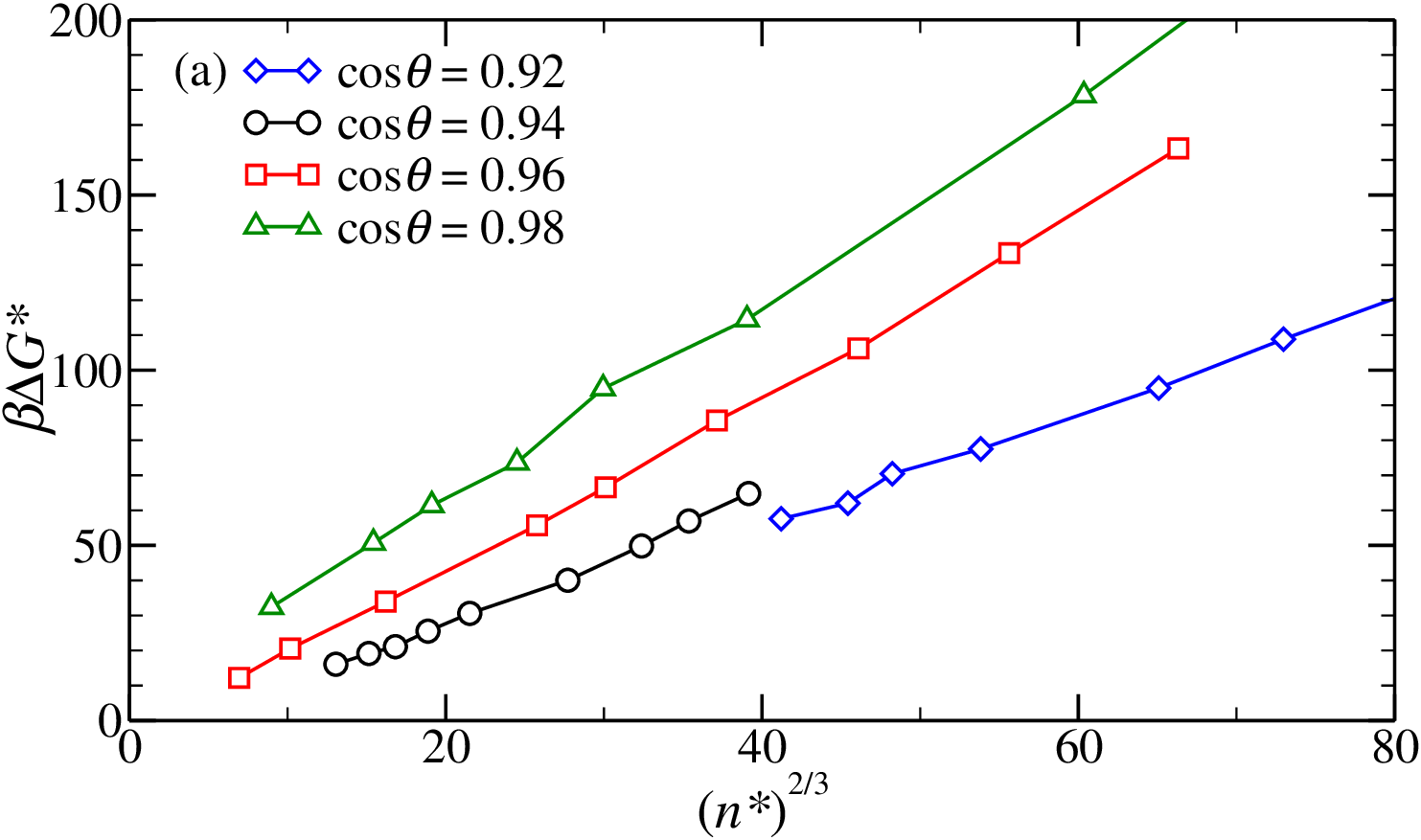}}
\hbox to \hsize{\epsfxsize=1\hsize\epsfbox{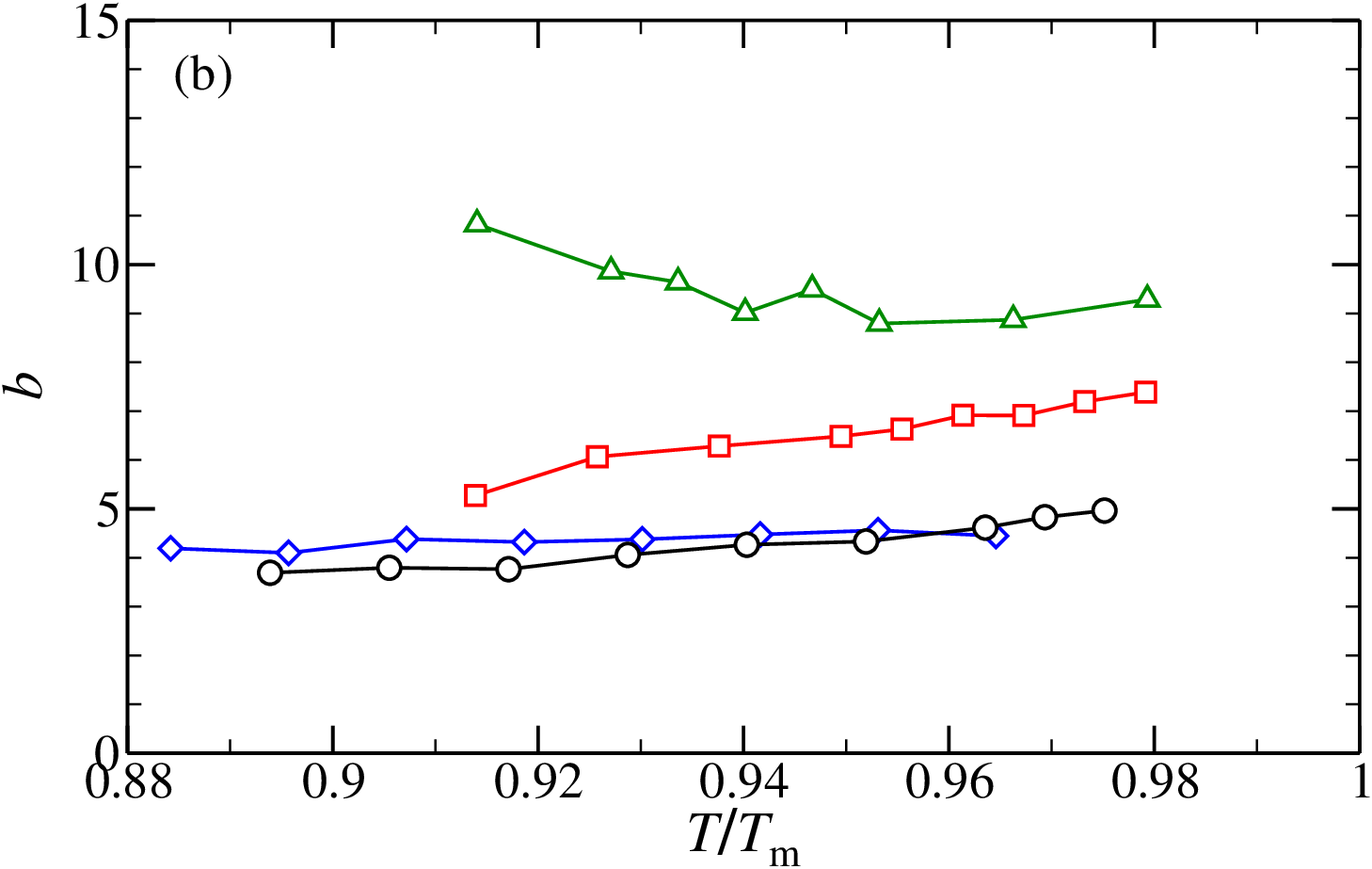}}
\hbox to \hsize{\epsfxsize=1\hsize\epsfbox{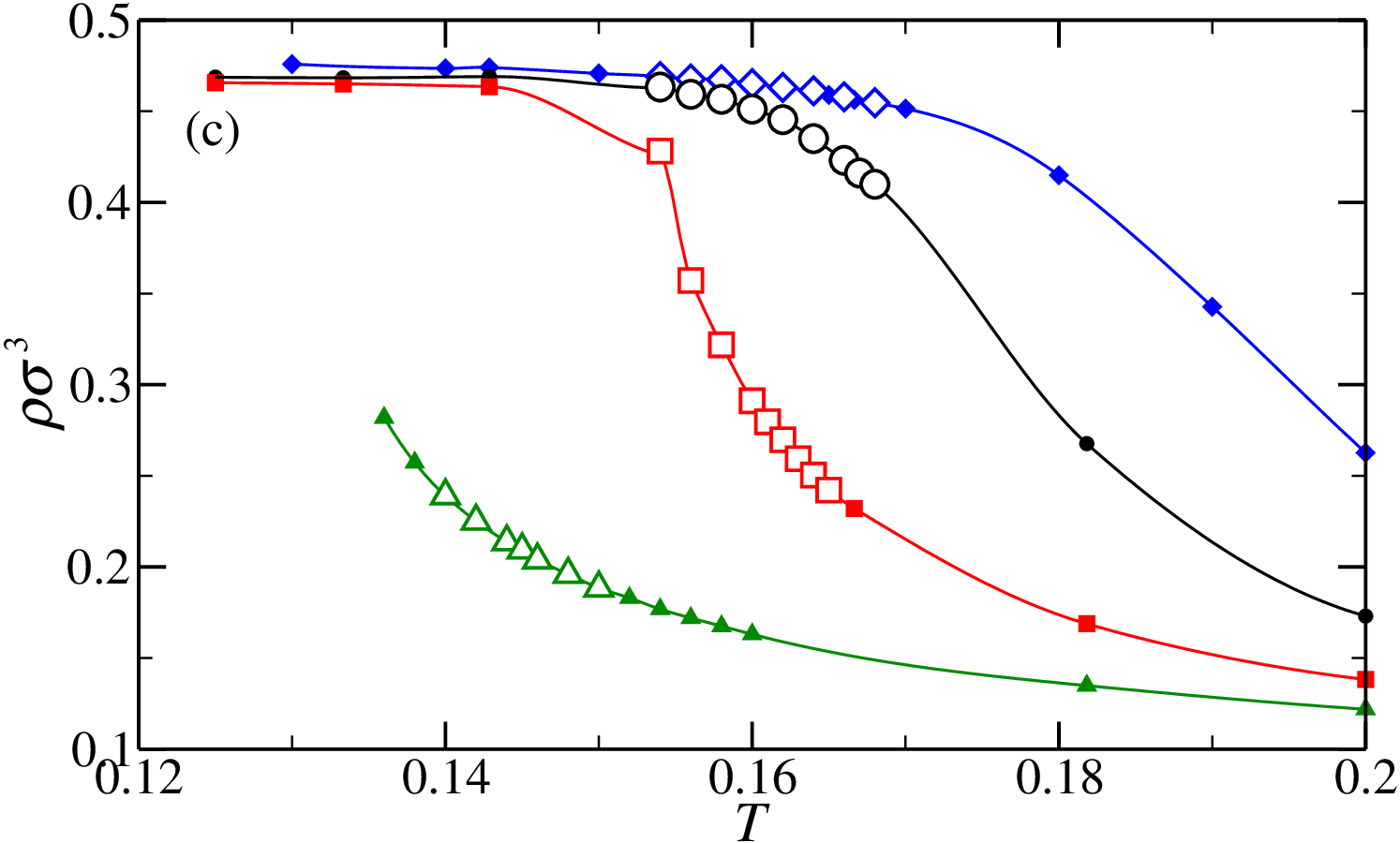}}
\caption{\label{fig:Nstar-DeltaG}
(a) Barrier height as a function of $n^{2/3}$ (proportional to the surface
area of the cluster). All models show a linear behaviour.
(b) Surface coefficient $b$ of the CNT fit as a function of the reduced
temperature. The surface term does not change significantly with $T$, as
compared to the barrier height [see Fig.~\ref{fig:gstar}(a)].
(c) Reduced number density $\rho \sigma^3$ as a function of $T$ at $P=0.03$
for the models studied. Large open symbols indicate the range of state
points used in the calculation of $b$. Curves are a guide to the eye.
}
\end{figure}

To illustrate the dependence of the surface tension on patch width,
motivated by the relation $3 \beta \Delta G^* = b n^{*2/3}$ that follows
from Eq.~\ref{eq:CNTfit}, we plot in Fig.~\ref{fig:Nstar-DeltaG}(a) $\beta
\Delta G^*$ as a function of $n^{*2/3}$. For all models, a reasonable
linear dependence is observed, with a slope that progressively increases on
with $\cos{\theta}$ beyond 0.94. This results in a larger critical size for
a given barrier height as the patch width increases. The near linearity
also suggests that the surface tension does not have a strong dependence on
supercooling. In Fig.~\ref{fig:Nstar-DeltaG}(b) we show $b$ as a
function of $T/T_m$ as obtained from the fitting procedure for the
different state points studied. The surface tension increases
significantly on decreasing the angular patch width. Once again, however,
the two models with the largest patches are quite similar in their
behavior.
 
It is worthwhile noting that the densities at our simulated $P$ for these
two wide patch models (at our lowest $T$) are very similar with values near
$\rho \sigma^3 = 0.47$, almost matching the crystal density (where $\rho$
is the number density). The density is smaller for $\cos{\theta}=0.96$, and
smallest for $\cos{\theta}=0.98$. Thus, the values of $b$ for the different
models appear to reflect the differences in density, with a better match in
density between fluid and crystal giving rise to a smaller surface tension.
However, a plot of the $T$ dependence of the densities in
Fig.~\ref{fig:Nstar-DeltaG}(c) shows that density does not solely determine
$b$, as indicated by the variation of density without a correspondingly
large variation in $b$ within each model.

\subsection{$a$ vs $\beta\Delta\mu$}

\begin{figure}[tb]
\hbox to \hsize{\epsfxsize=1\hsize\epsfbox{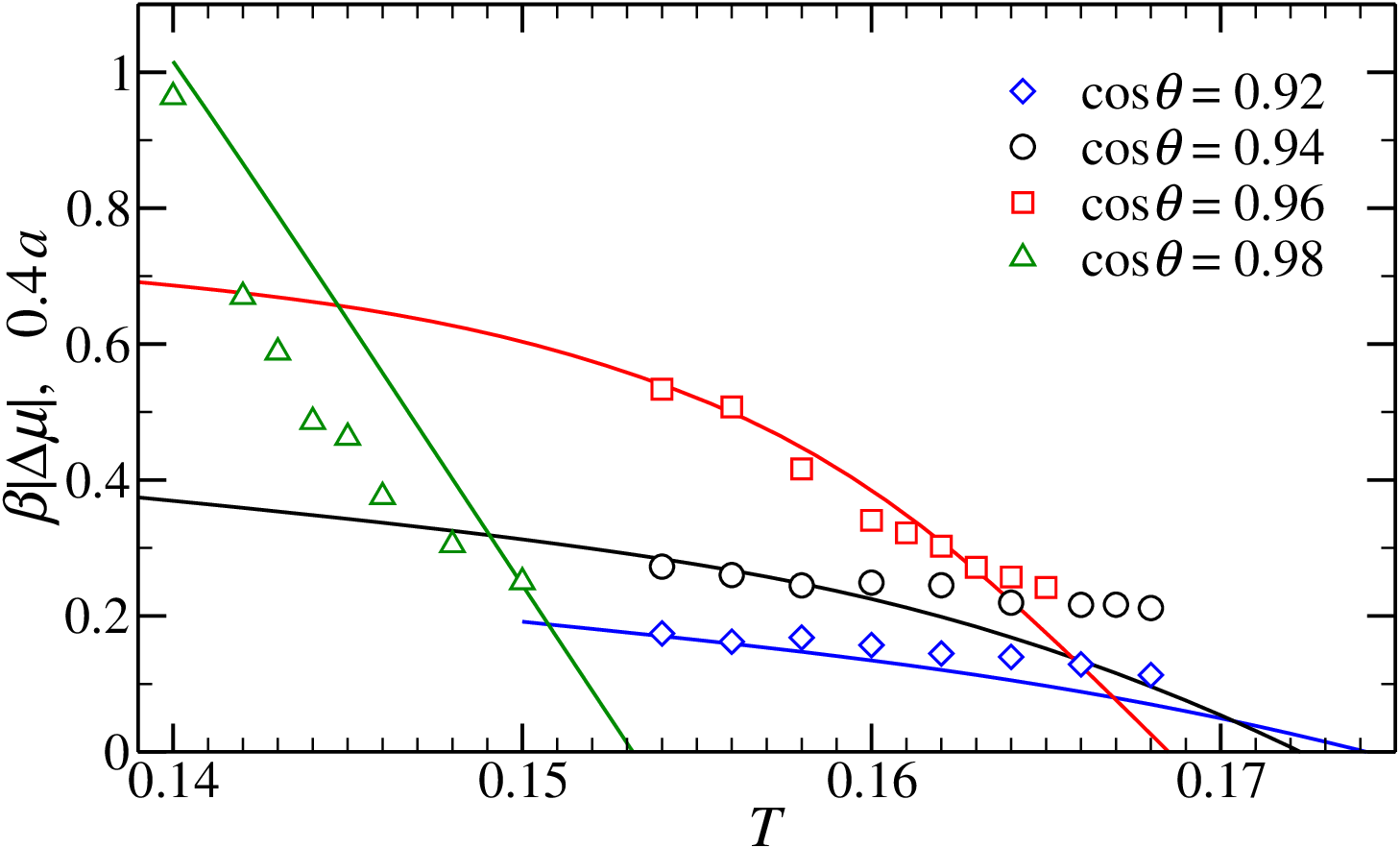}}
\caption{
\label{fig:Mu-T-compare}
CNT bulk fit parameter $a$ as a function of temperature (points) compared
with the difference in chemical potential $\beta \left | \Delta \mu \right |$
(lines). While the two quantities show similar trends, to achieve
quantitative agreement we multiply values of $a$ in the plot by a universal
scaling factor of 0.4.
}
\end{figure}

In Fig.~\ref{fig:Mu-T-compare} we plot $\beta |\Delta \mu|$~\cite{melting}
and $a$ (obtained from the CNT fits) as functions of $T$. For comparison
purposes, all the values of $a$ appearing in the figure have been
multiplied by a factor of 0.4, i.e. the values of $a$ are
significantly higher than expected from independent calculation of $\beta |
\Delta \mu|$. However, it is quite comforting that the rescaled $a$ matches
$\beta \Delta \mu$ quite well across $T$ and $\theta$, as this strengthens
our assumption that the fit parameter $b$ is proportional to $\beta \gamma$
with the same proportionality constant across all the models.

On the other hand, it is somewhat perplexing that such a large rescaling is
required at all. In a few previous studies, comparison between $a$ and
$\beta \Delta \mu$ have shown that in some cases a close correspondence is
observed~\cite{frenkel_2004, frenkel_2005, SaikaBKS}, while in principle the value 
of  $a$ should vary with the definition crystal-like particles. 
There are several potential reasons why
such a disagreement can be found. First of all, the assumption of CNT may
not be fully satisfied, including those concerning the structure of the crystallites and the sharpness of the
interface~\cite{lechner}.
Second, an imperfect classification of particles
as solid-like leads to an imperfect evaluation of the cluster size and a
classification-dependence of $a$ and $b$.

\begin{figure}[tb]
\hbox to \hsize{\epsfxsize=1\hsize\hfil\epsfbox{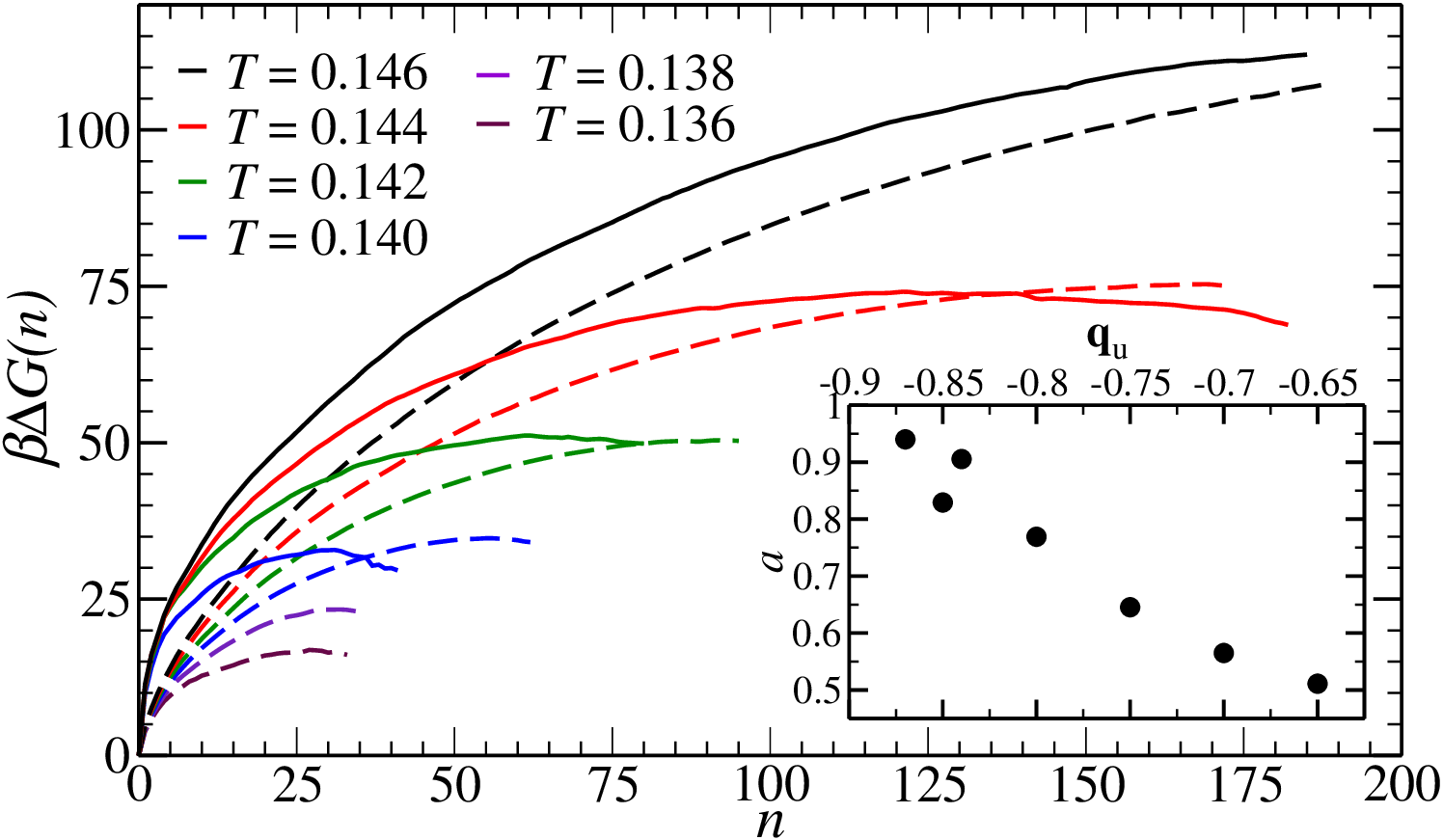}}
\caption{
\label{fig:q65q87}
Comparison of the nucleation barriers obtained with two different order
parameters, $q_u=- 0.65$ (dashed lines) and $q_u=
-0.87$ for the $\cos{\theta}=0.98$ model. As expected, the barrier heights are consistent (see
Ref.~\cite{Filion2010}) but the critical cluster sizes differ significantly.
The more relaxed order parameter leads to a larger critical cluster, possibly
due to the fact that it includes more particles on the surface. In the inset
we also show the dependence of the bulk CNT term $a$ as a function of the order
parameter, showing that it increases by making the order parameters more
strict.
}
\end{figure}

As a test of this second hypothesis and to illustrate the effect of the
parameters chosen to define crystalline clusters, we plot in
Fig.~\ref{fig:q65q87} sets of barrier profiles obtained for
$\cos{\theta}=0.98$ using $q_u=-0.65$ compared against using $q_u=-0.87$.
While $\Delta G^*$ remains invariant to within error with respect to $q_u$,
at all investigated $T$, $n^*$ increases, as expected, for smaller values
of $q_u$. In other words, the size of the critical nucleus is strongly
dependent on the criteria chosen to define solid-like particles. It is not
a surprise that $\Delta G^*$ is a much more robust value, since the work
required to produce the critical nucleus is independent on how the cluster
is described. In other words, the ``real'' critical nucleus (i.e.
configurations which 50 per cent of the time will crystallize and 50 per
cent will melt again into a fluid state) is what needs to be sampled in the
simulation. The number of particles that compose this cluster depends on
the definition, affecting the perceived size, but not the barrier height.
These conclusions are in agreement with the work of Filion {\it et
al}~\cite{Filion2010} on spherical particles.

In the inset of Fig.~\ref{fig:q65q87}, we show $a$ as a function of $q_u$
for $T=0.146$ and $\cos{\theta}=0.98$ and see that $a$ varies by roughly a
factor of 2 for the reasonable range of $q_u$ we have explored.
Coincidentally, the value for $\beta |\Delta \mu|$ for this state point is
0.56. Thus, for an appropriate choice of $q_u$, Eqs.~\ref{CNT} and
\ref{eq:CNTfit} describe the nucleation barrier profiles both in scaling
and in the numerical value of $\Delta \mu$, but how to choose the
appropriate $q_u$ {\it a priori} is not clear. 

The more inclusive choice of $q_u=-0.65$ allows one to more easily explore the
regime of low barrier heights, less hampered by the formation of secondary
clusters. Using $q_u=-0.65$ we are able to add two additional points to the
curve for $\cos{\theta}=0.98$ in Fig.~\ref{fig:gstar}(a).

\subsection{DH vs DC}

\begin{figure}[tb]
\hbox to \hsize{\epsfxsize=1\hsize\hfil\epsfbox{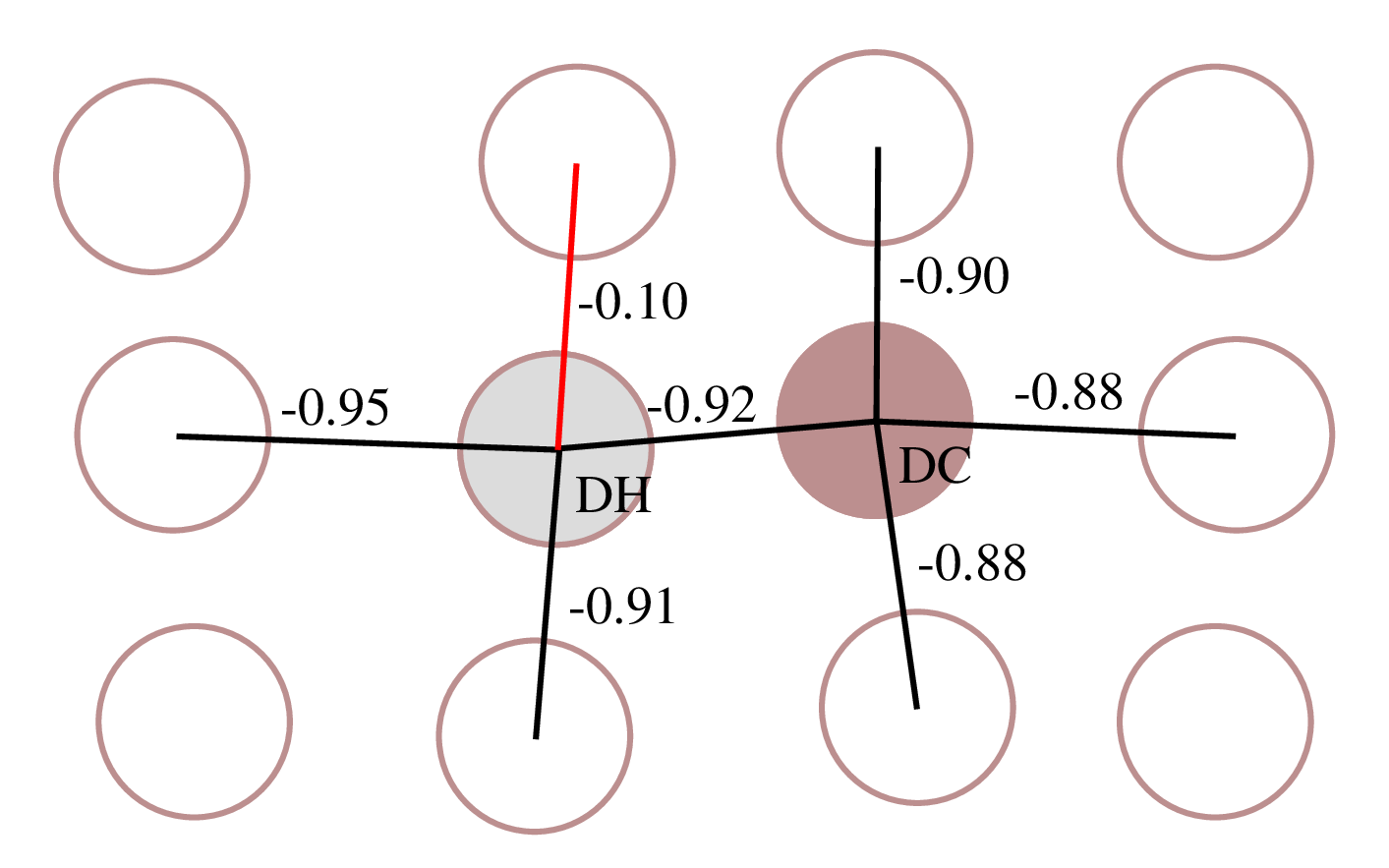}}
\caption{
\label{fig:cartoon}
Schematic description of the criteria that require four connections for
defining solid-like particles. The values of $c_{ij}$ are displayed next to
the bonds that the two central particles form. The particle on the left has
three bonds with $c_{ij}<-0.87$ and one bond with $-0.3 <c_{ij}<0.1$,
registering it as a DH particle. The particle on the right has four bonds
with $c_{ij}<-0.87$, registering it as a DC particle. In the mixed
four-connections case, both particles would be considered solid-like.
}
\end{figure}

\begin{figure}[tb]
\hbox to \hsize{\epsfxsize=1\hsize\hfil\epsfbox{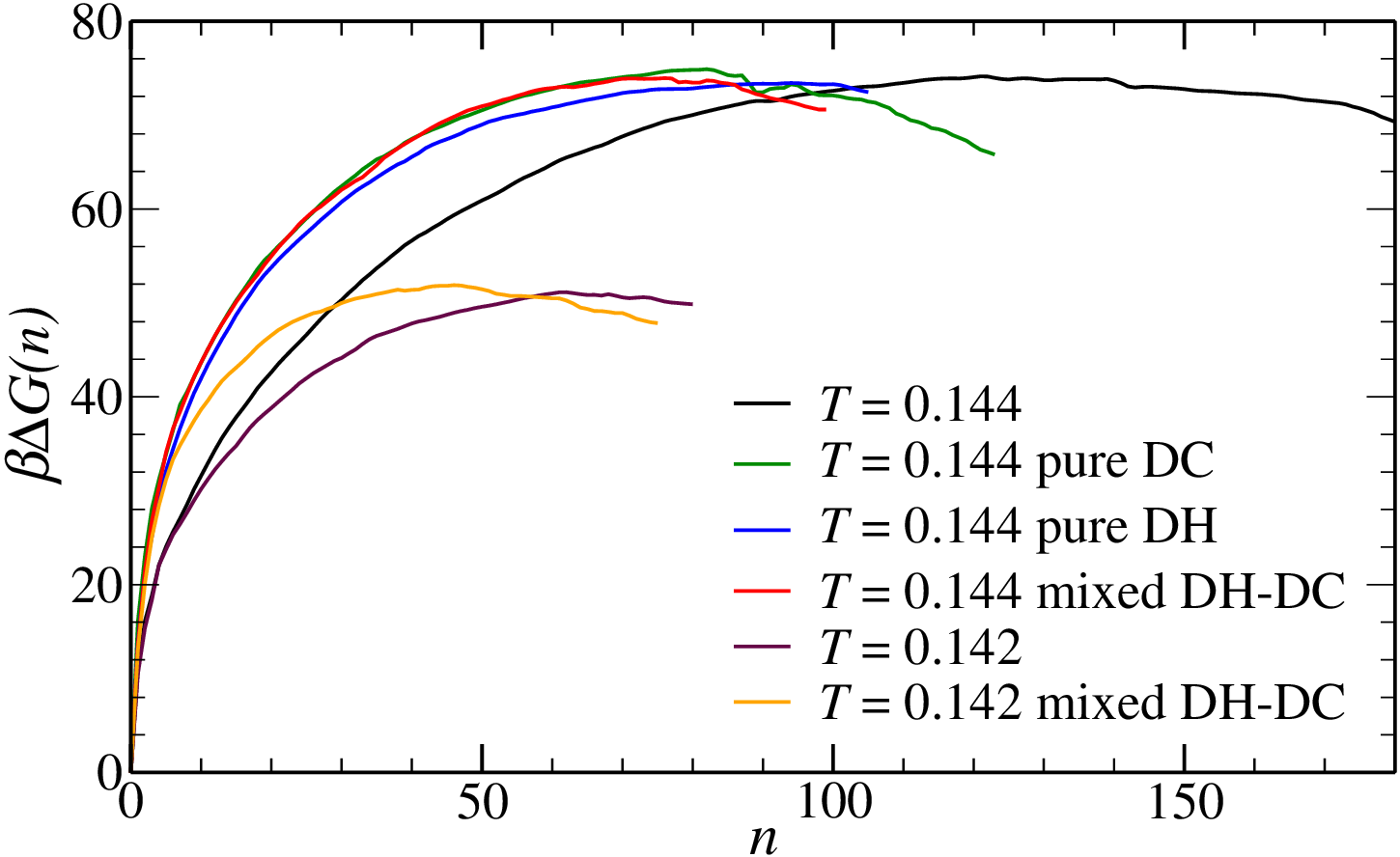}}
\caption{
\label{fig:dhdc}
Nucleation barriers for pure DC, pure DH, and mixtures of the two. For
$T=0.144$ the four-connections curves (pure DC, pure DH and mixed DH-DC)
are all quite similar to each other and yield the same barrier height as
the three-connections case, but smaller $n^*$. Barriers at $T=0.142$
compare the mixed three- and four-connections cases and confirm the
invariance of $\Delta G^*$ and the change in $n^*$ with different 
criteria for defining clusters.
}
\end{figure}

As we have alluded to before, the clusters are formed by a mixture of local
DC and DH particles. Indeed, the criterion of having at least three
connected neighbors ($c_{ij}< q_u$) for defining solid-like particles,
accepts either a locally DC or DH particle as solid. However, this
definition is more likely to misidentify a DH-like particle as liquid-like,
since particles in the DC crystal have four such connections, while DH
particles only have three. The biasing potential may therefore tend to
grow clusters richer in DC than would occur naturally. In the following we
therefore compare this criterion with other criteria which enforce
the growth of pure DC, pure DH or an unbiased mixture of the two. We do
this at $\cos{\theta}=0.98$, where the calculations are the fastest.

For pure DC, we retain $q_u=-0.87$ and simply require that the number of
connections be exactly four (as opposed to at least three). With this more
restrictive criterion, the solid-like particles must have a DC environment.
For pure DH, we require exactly three connections with $c_{ij} < q_u =
-0.87$ and one connection with $-0.3 < c_{ij} < 0.1$. In this way, DC
particles are excluded and can not participate in crystalline clusters.
For the mixed DH/DC case, a particle is solid-like if it is determined
again by exactly four connections, but this time these four connection must
satisfy either the criterion for a DC particle or the criterion for a DH
particle. The cartoon in Fig.~\ref{fig:cartoon} provides a graphical
explanation of the criteria for identifying solid-like particles.

In Fig.~\ref{fig:dhdc}, we show the barrier profile results at $T=0.144$
with $\cos{\theta}=0.98$ for the four different definitions of solid-like
particles (three-connections, pure DC, pure DH and unbiased
four-connections mixture), in order to asses if the energy barriers of the
pure crystals are different from that of the mixture and to ascertain the
difference between using three or four connections in the mixed case. We
can not discern any real difference in terms of the barrier height between
the four criteria. This interesting observation supports the view for
this short-range model, that
in addition to the DH and DC free energies being
essentially identical, the surface tensions are also similar.
%
Even more, if we restrict ourselves to the comparison of all criteria
in which four connections are required, then not only is the barrier height
identical, but also (within our numerical precision) the $n$ dependence of
the barrier profile. 
This suggests that the pathways to crystallizing DC and DH are quite similar as well.
The pure DH case appears to produce a slightly larger
$n^*$, but this will also likely depend on the bounds on $c_{ij}$ near
$-0.1$. Comparing the three-connections criterion with the more
restrictive four-connections cases, we see that the four-connections
criterion simply results in an apparently smaller critical cluster size,
providing one additional piece of evidence for the sensitivity of the
profile on the solid-like definition accompanied by the insensitivity of
the barrier height.
 
We repeat the comparison at the lower $T=0.142$. However, we are unable to
construct barriers for the pure DC and DH cases. At this temperature and
even at moderate values of $n_0$, the presence of a pure DC
cluster appears to act as a template for DH growth (and viceversa).
Essentially, the DH particles, which are not counted as solid-like,
become part of the growing crystal, providing a bridge between only
apparently different DC crystal clusters. Thus, the system crystallizes
while still registering a small largest cluster in the system and our order
parameter no longer describes nucleation.

This issue is greatly reduced for the four-connection mixed case and we are
able to calculate the barrier at $T=0.142$. Again, the barrier height is
comparable to the one calculated for the case of three connections. At
lower $T$, also for the four-connections mixed case, it becomes impossible
to properly evaluate the barrier, since  the
overly restrictive criterion will misidentify as liquid-like particles
sufficiently crystalline to participate in crystal growth.

While the barrier heights are the same across the different cluster
criteria, the DC/DH composition of the clusters varies. To quantify this in
a basic way, we analyze an ensemble of largest clusters extracted from a
set of US simulations employing the mixed four-connections solid-like
criterion in a wide range of $n_0$ values. We find that 54\% of the solid
particles are DC (standard deviation 17), which is the same as the
composition for clusters appearing in spontaneous nucleation studies as
reported in Ref.~\cite{FLAVIOXstal}. Performing the same evaluation
starting from US simulations, employing this time the mixed
three-connections solid-like criterion ($q_u=-0.87$), the fraction of DC
particles within the largest cluster is 73\% (standard deviation 22). The
result is the same for $q_u=-0.65$. This confirms that the choice of at
least three connections with $c_{ij}$ close to minus one does indeed
introduce a bias toward the DC structure.
For this model, however, this bias does not measurably affect the nucleation barrier.

\subsection{BCC}

For all the patch widths considered except $\cos{\theta}=0.98$ we are
working in the stability field for DC/DH. For the case of
$\cos{\theta}=0.98$, the stable crystal phase at $P=0.03$ is the BCC (see
Fig.~\ref{fig:phds-barriers}), even if spontaneous nucleation indicates
that the crystal that forms is the DC/DH mixture. To verify that the
barrier for nucleating the BCC is significantly larger than the one for
nucleating DC or DH or their mixture for $\cos{\theta}=0.98$, we evaluate
the barrier at $T=0.142$ using the same procedure described for the
tetrahedral crystals. The definition of solid-like BCC particle is based on
the $l=6$ spherical harmonics~\cite{frenkel_2004}, defining neighbors to be
connected when the scalar product of the $l=6$ harmonics is greater than
0.5. A particle is classified as a solid-like if it has at least six
connections. We note that Ref.~\cite{frenkel_2004} does not use normalized
$q_{lm}(i)$ vectors for calculating $c_{ij}$, so we have determined cutoffs
based on our distributions of $c_{ij}$ and of the subsequent number of
connections a particle has in the liquid and in the crystal. We do find
that the barrier to nucleating BCC is at least $70\,k_{\rm B}T$ larger than for
tetrahedral crystals, supporting the lack of observation of spontaneous BCC
nucleation at the pressure we consider here.

\section{Summary and Conclusions}

Previous work for this tetrahedral patchy model showed that the driving
force for nucleation (quantified as difference in the crystal and fluid
chemical potentials) decreases as the patch width increases.
Fig.~\ref{fig:Mu-T-compare} summarizes this result by showing $\beta
|\Delta \mu|$ increasing rapidly to large values as $T$ decreases below
$T_m$ for narrow patches, while increasing only slowly and remaining fairly
small for wide patches.

We find that despite this reduced driving force, the barriers to nucleation
actually are smallest for the $\cos{\theta}=0.96$ case [as shown in
Fig.~\ref{fig:gstar}(a)], in the sense that this model reaches the
homogeneous nucleation limit, marked for example by the barrier reaching
a value of $\beta \Delta G^*=10$, at the highest $T/T_m$ amongst the
studied models. The increased similarity between liquid and crystal in
terms of energy and density as patch width increases not only brings the
chemical potentials of liquid and crystal closer in value (tending to
increase the nucleation barrier), but also reduces the surface tension
(tending to lower the barrier). Thus, in the range of narrow angles where 
crystallization is readily observed,
competition between $| \Delta \mu|$
and $\gamma$ leads to optimal nucleation at an intermediate patch width. 

\begin{figure}[tb]
\hbox to \hsize{\epsfxsize=1\hsize\hfil\epsfbox{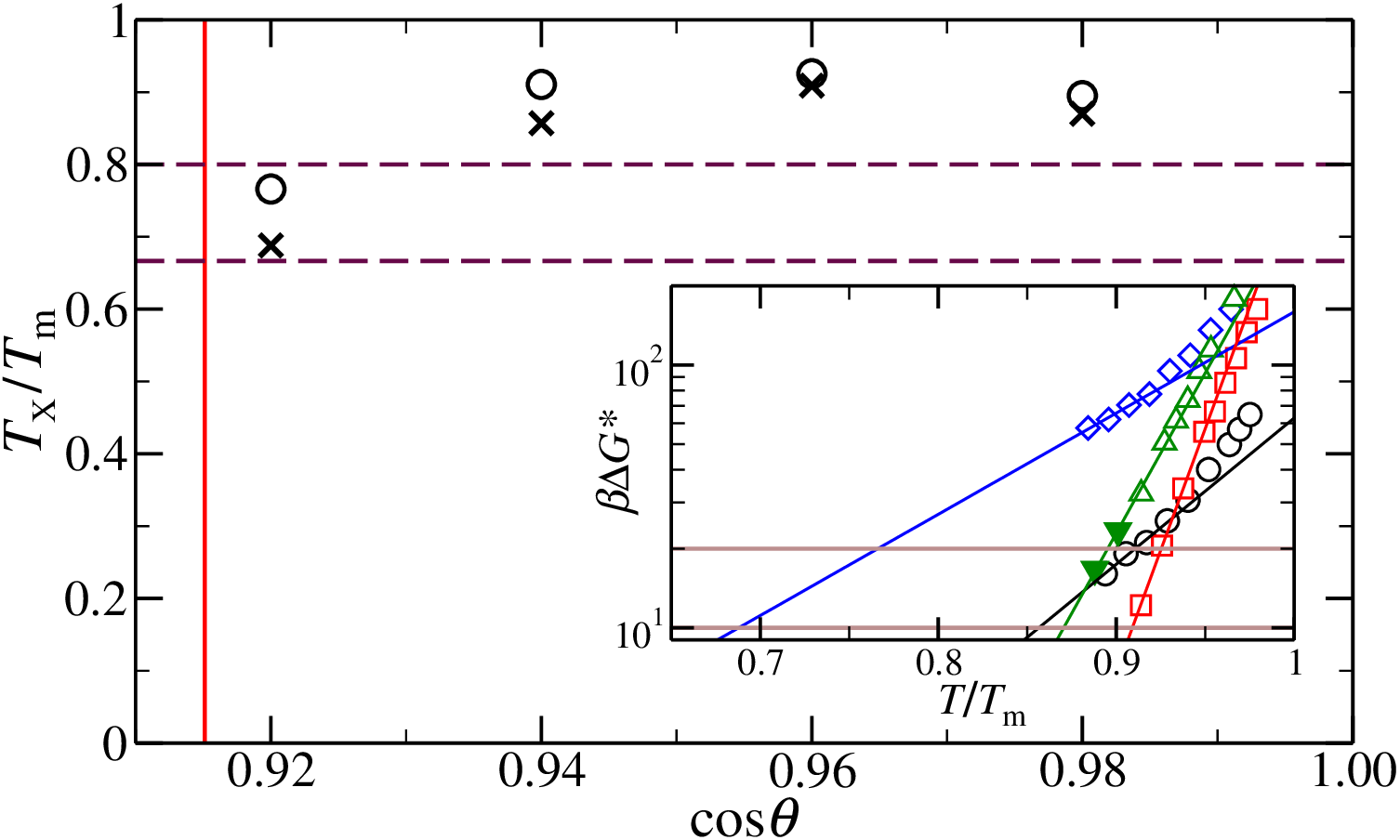}}
\caption{
\label{fig:T10}
Estimates of homogeneous nucleation temperature $T_{\rm X}^{10}$ (crosses)
and $T_{\rm X}^{20}$ (circles) as functions of $\cos{\theta}$. The solid
vertical line indicates the value of $\cos{\theta}=0.9151$ at which more
than one bond per patch becomes possible for $\delta=0.24$. The upper
horizontal dashed line marks the approximate value of $T_g/T_{\rm m}=0.8$
for silica, while the lower one indicates the general value of $T_g=2T_{\rm
m}/3$. Inset shows the determination of $T_{\rm X}^{10}$ and $T_{\rm
X}^{20}$ from the crossing of exponential fits to the four lowest $T$
points for each value of $\cos{\theta}$ with $\beta \Delta G^*=10$ and
$20$, respectively.
}
\end{figure}

Increasing the patch width beyond $\cos{\theta}=0.94$ no longer
significantly reduces $\gamma$, while $| \Delta \mu|$ continues to
decrease, causing an increase in nucleation barrier heights. For
$\cos{\theta}=0.92$, this resulting increase is quite large, with $\beta
\Delta G^*$ estimated to be of order 50 at the lowest $T$ that we can simulate.
Moreover, the rate of decrease of $\beta \Delta G^*$ with $T$ appears to be
quite slow, requiring significant supercooling to reach accessible
nucleation barrier heights.

The evaluation of the barriers requires a definition of solid-like
particles. We have checked that the barrier height is essentially
insensitive to the exact choice of the cutoff used to define solid-like
connections, consistent with the observation in Ref.~\cite{Filion2010} for
hard-spheres. The size of the critical nucleus is instead significantly
dependent on the definition of solid-like particles, again in agreement
with the conclusions in Ref.~\cite{Filion2010}. We have also compared
several definitions of solid-like particles to probe the crystallization of
a pure DC, a pure DH and a mixed DC-DH structure. Interestingly, we find
the same value of $\beta \Delta G^*$ for all structures and, in this case,
similar $n^*$. The fact that the entropy gain in creating a mixed
structure does not appreciable lower the work of forming a critical
nucleus, but is sufficient to generate a prevalence of mixed structures
when spontaneous nucleation takes place~\cite{FLAVIOXstal}, warrants
further investigation. 

At this point, we would be remiss if we were not to comment on the
interplay between dynamics and thermodynamics in controlling the rate of
nucleation, and indeed governing the glass-forming abilities of the system.
We already see an example of this within our data. The barriers at the
lowest $T$ studied for $\cos{\theta}=0.98$ and $\cos{\theta}=0.94$ are the
same within error, and have a value of $\beta \Delta G^*\approx 17$.
However, as mentioned previously, the dynamics (in terms of the diffusion
coefficient) are 40 times slower for the wider patch case. For
simulations, this is a significant number. The slow decrease in $\beta
\Delta G^*$ with $T$, combined with the expected slow-down in dynamics,
suggest that a search for spontaneous nucleation in unconstrained
simulations would target $T$ not far from the lowest $T$ for which we have
calculated the barrier. The case of $\cos{\theta}=0.92$ is more obvious
since its dynamics are even slower at a given $T$ than those of
$\cos{\theta}=0.94$: there is no hope of seeing nucleation in this model,
which for all intents and purposes is a glass-former.

To place these results in slightly broader context, that may be
illuminating for molecular tetrahedral liquids that are quite different in
their glass-forming and crystallizing properties, we use the nearly
exponential behavior of $\beta \Delta G^*$ to find estimates for the temperature at which the homogeneous nucleation limit is reached, $T_{\rm X}$.  We define two such estimates using  $\beta \Delta G(T_{\rm X}^{10})=10$ and $\beta \Delta G(T_{\rm X}^{20})=20$, and determine them by 
fitting the lowest four points in $T$ from Fig.~\ref{fig:gstar}(a)
for each model to exponentials and extrapolate where necessary.
In Fig.~\ref{fig:T10} we present the resulting estimates of $T_{\rm X}$
alongside a vertical line showing the widest patch that guarantees a single
bond per patch, i.e. four bonds per particle. Also shown are two horizontal
lines indicating the approximate value of $T_g/T_{\rm m}=0.80$ for
silica~\cite{Tg-silica} (where $T_g$ is the glass transition temperature),
as well as the general rule of thumb $T_g/T_{\rm
m}=2/3$~\cite{DebenedettiGlass} which has been shown to be valid for a
large class of molecular~\cite{Tg-inorganic} and polymeric
systems~\cite{Tg-polymers}. For $T_{\rm X}/T_{\rm m}$ below $T_g/T_{\rm
m}=2/3$, a model may be considered a glass-former; our estimates suggest
that we may be approaching this regime with our widest patch model. The
presence of a maximum in $T_{\rm X}$ confirms that decreasing the angular
range for bonding favors glass formation. Tetrahedral colloidal particles
will thus form crystals only if the bonding angular width is small.

As a final remark, we point out that our findings may shed some light on the glass-forming
and crystallizing abilities of molecular or atomic tetrahedral network-forming liquids,
contributing insight as to why, for
example, water crystallizes while silica more readily forms a glass.

\section{Acknowledgments}
We thank ACEnet (Canada) for computational resources.
I.S.-V. acknowledges financial support from NSERC (Canada). FS and FR acknowledge support from
ERC-226207-PATCHYCOLLOIDS and ITN-COMPLOIDS.


\begin{thebibliography}{999}
\bibitem{melting}
Here, $\Delta \mu$ refers specifically to the difference in chemical potential between the DC crystal and the fluid,
as calculated in Ref.~\cite{FLAVIOXstal}, even in the case of $\cos{\theta}=0.98$, where the stable phase is BCC.
$T_m$ is determined from the condition $\Delta \mu=0$.
\end{thebibliography}
\end{document}